\documentstyle[aps, array, manuscript]{revtex}
\tightenlines
\draft
\begin{document}
\title{Inflaton and metric fluctuations in the early universe
from a 5D vacuum state}
\author{
$^{1}$ Agustin Membiela\footnote{
E-mail address: membiela@argentina.com}
and $^{1,2}$Mauricio Bellini\footnote{
E-mail address: mbellini@mdp.edu.ar}}
\address{
$^1$Departamento de F\'{\i}sica,
Facultad de Ciencias Exactas y Naturales,
Universidad Nacional de Mar del Plata,
Funes 3350, (7600) Mar del Plata, Argentina.\\
$^2$ Consejo Nacional de Ciencia y Tecnolog\'{\i}a (CONICET).}

\vskip .5cm
\maketitle
\begin{abstract}
In this letter we complete a previously introduced formalism to
study the gauge-invariant metric fluctuations from a noncompact
Kaluza-Klein theory of gravity, to study the evolution of the early
universe. The evolution of both, metric and inflaton field fluctuations
are reciprocally related. We obtain that
$\left<\delta\rho\right>/\rho_b$
depends on the coupling of $\Phi$ with $\delta\varphi$ and
the spectral index of its spectrum is $0.9483 < n_1 < 1$.
\end{abstract}
\vskip .2cm
\noindent
Pacs numbers: 04.20.Jb, 11.10.kk, 98.80.Cq \\
\vskip 1cm
\section{Introduction and motivation}

The key property of the laws of physics that makes inflation possible
is the existence of states of matter that have a high-energy density
which cannot be rapidly lowered. In the original version of the
inflationary theory\cite{Guth}, the proposed state was a scalar field in
a local minimum of its potential energy function. A similar proposal was
advanced by Starobinsky\cite{Sta}, in which the high-energy density state
was achieved by curved space corrections to the energy-momentum tensor
of a scalar field. The scalar field state employed in the original version
of inflation is called a false vacuum, since the state temporarily acts as if
it were the state of lowest possible energy density. Chaotic inflation
is driven by a scalar field called inflaton, which in its standard
version, is related to a potential with a local minimum or a gently
plateau\cite{lin}. In this relativistic theory, cosmological perturbations
are a cornerstone in our understanding of the early universe and are
indispensable in relating early universe scenarios\cite{Muk}.

Higher dimensional spacetime is now an active field of activity in both
general relativity and particle physics in its attempts to unify
gravity with all other forces of nature\cite{cha}. Higher dimensional
unified-field theories include
Kaluza-Klein, induced matter, super string supergravity and
string theory.
In these $(4+d)$-dimensional models the $d$-spacelike dimensions
are generally spontaneously compactified and the symmetries of this space
appear as gauge symmetries of the effective 4D theory. In higher-dimensional
theories of gravity with large extra dimensions, the cylinder
condition of the old Kaluza-Klein theory is replaced
by the conjecture that the ordinary matter and
fields are confined to a 4D subspace usually refered to as
``3-brane''\cite{arkani}. Randall and Sundrum showed, for $d=1$, that
there is no contradiction between Newton's $1/r^2$ law of gravity
in 4D and the existence of more than 4 noncompact dimensions if the
background metric is nonfactorizable\cite{randall}. This has motivated
a great interest in brane-world models where our 4D universe
is embedded in a 5D noncompact spacetime\cite{otros}. On the other
hand, another noncompact theory is the so called space-time-matter (STM)
theory. In STM the conjecture is that the ordinary matter and fields
that we observe in 4D result from the geometry of the extra
dimension\cite{we}.

This paper is devoted to study a 4D de Sitter expansion of the universe
from a noncompact Kaluza-Klein (NKK)
theory of gravity, taking into account in a consistent
manner both, the gauge-invariant scalar metric and inflaton field
fluctuations. It was done, but in a partial manner, in a previous
work\cite{plb2006}.

\section{Review and extension of the Formalism}

In the framework of a NKK theory we shall consider
an action for a scalar field $\varphi$, which
is minimally coupled to gravity on a 5D manifold 
\begin{equation} \label{action}
I=-\int d^{4}x \  d\psi\,\sqrt{\left|\frac{^{(5)}
\bar g}{^{(5)}\bar g_0}\right|} \ \left[
\frac{^{(5)}\bar R}{16\pi G}+ ^{(5)}{\cal 
L}(\varphi,\varphi_{,A})\right],
\end{equation}
where $G=M^{-2}_p$ is the gravitational constant
and $M_p=1.2 \  10^{19} \  {\rm GeV}$ is the Planckian mass.
To describe a manifold in apparent vacuum we shall consider a
Lagrangian density ${\cal L}$ in (\ref{action}), which is only
kinetic
\begin{equation}
^{(5)} {\cal L}(\varphi,\varphi_{,A}) = \frac{1}{2} g^{AB}
\varphi_{,A} \varphi_{,B}.
\end{equation}
Here, $A,B$ can take the values $0,1,2,3,4$ 
and the Ricci scalar
$^{(5)}\bar R=0$ in (\ref{action})
is evaluated on the background metric\cite{diego}
\begin{equation}\label{back}
\left(dS^2\right)_b
= \psi^2 dN^2 - \psi^2 e^{2N} dr^2
-d\psi^2,
\end{equation}
which is 3D spatially isotropic, homogeneous and flat\cite{otro}.
Furthermore, it is globally flat (i.e.,$\bar R^A_{BCD} =0$).
Here, $N,x,y,z$ are dimensionless and $\psi$ has spatial units.
This background metric
describes an apparent
vacuum, because $\bar G_{AB}=0$.
For the metric (\ref{back}), $|^{(5)}\bar g|=\psi^8 e^{6N} $
is the absolute value for the determinant of $\bar g_{AB}$ 
and
$|^{(5)} \bar g_0|=|^{(5)}\bar g|_{N=N_0,\psi=\psi_0}= \psi^8_0 e^{6N_0}$
is a dimensional constant.
In this work we shall consider $N_0=0$, so that
$\left|^{(5)}\bar g_0\right|=\psi^8_0$.

In order to describe scalar metric fluctuations we
must consider a symmetric energy-momentum
tensor. In a longitudinal gauge the perturbed line element
is given by\cite{plb2006}
\begin{equation}   \label{1}
dS^2
= \psi^2 \left(1+ 2 \Phi\right) dN^2 - \psi^2
\left(1- 2 \Phi\right) e^{2N} dr^2 - \left(1- 2 \Phi\right) d\psi^2,
\end{equation}
where the field $\Phi(N, \vec r, \psi)$ is the scalar metric perturbation
of the background 5D metric (\ref{back}).

The contravariant metric tensor, after
a $\Phi$-first order approximation, is given by
\begin{displaymath}
g^{AB} = {\rm diag}\left[ (1-2\Phi)/\psi^2, -(1+2\Phi)e^{2N}/\psi^2,
-(1+2\Phi)e^{2N}/\psi^2, -(1+2\Phi)e^{2N}/\psi^2, -(1-2\Phi)\right],
\end{displaymath}
which can be written as $g^{AB} = \bar g^{AB} + \delta g^{AB}$, being
$\bar g^{AB}$ the contravariant background metric tensor.

If we use the semiclassical approximation $\varphi(N,\vec r,\psi) =
\varphi_b(N,\psi) + \delta\varphi(N,\vec r,\psi)$, the Lagrange equations
(we use a linear approximation on $\delta\varphi$
and $\Phi$) for $\varphi_b$ and
$\delta\varphi$ are
\begin{eqnarray}
&& \frac{\partial^2\varphi_b}{\partial N^2}
+ 3 \frac{\partial\varphi_b}{\partial N}
-\psi \left[ \psi \frac{\partial^2\varphi_b}{\partial\psi^2} +
4 \frac{\partial\varphi_b}{\partial\psi} \right]=0, \label{l3} \\
&& \frac{\partial^2\delta\varphi}{\partial N^2}
+ 3 \frac{\partial\delta\varphi}{\partial N}
- e^{-2N} \nabla^2_r \delta\varphi - \psi \left[
4\frac{\partial\delta\varphi}{\partial\psi}
+ \psi \frac{\partial^2\delta\varphi}{\partial\psi^2} \right] \nonumber \\
& - & 2 \frac{\partial\varphi_b}{\partial N} \frac{\partial\Phi}{\partial N}
-2 \psi^2 \left[ \frac{\partial\varphi_b}{\partial\psi}
\frac{\partial\Phi}{\partial\psi}
+\left( \frac{\partial^2\varphi_b}{\partial\psi^2}
+\frac{4}{\psi} \frac{\partial\varphi_b}{\partial\psi}\right)
\Phi\right] =0, \label{l4} 
\end{eqnarray}
where $\varphi_b$ is the solution of equation of motion for
the inflaton field in absence of the inflaton
and metric fluctuations [i.e., for $\Phi=\delta\varphi=0$].
On the other hand, one obtains\cite{plb2006}
\begin{equation}\label{l5}
\left(\frac{\partial\varphi_b}{\partial N}\right)^2 + \psi^2
\left(\frac{\partial\varphi_b}{\partial\psi}\right)^2 =0.
\end{equation}
From the diagonal first order Einstein's equations $\delta G_{AA} = -8\pi G
T_{AA}$, and taking into account the eq. (\ref{l5}), we obtain
\begin{equation} \label{phi}
\frac{\partial^2 \Phi}{\partial N^2}+ 3 \frac{\partial\Phi}{\partial N}
- e^{-2N} \nabla^2_r \Phi - 2 \psi^2 \frac{\partial^2\Phi}{\partial\psi^2}
=0.
\end{equation}
This equation was obtained in \cite{plb2006},
and describes the evolution for the 5D scalar metric fluctuations
$\Phi(N,\vec r, \psi)$.\\

Once we know $\Phi$ from eq. (\ref{phi}) we must calculate the
inflaton field fluctuations from the equation (\ref{l4}). To do
it, firstly, we must solve the equation (\ref{l3}) for the
background inflaton field $\varphi_b$. If we propose
that $\varphi_b(N,\psi) = \varphi_1(N) \varphi_2(\psi)$, we obtain
the following equations
\begin{equation}\label{sep}
\frac{1}{\varphi_1} \frac{\partial^2\varphi_1}{\partial N^2}
+\frac{3}{\varphi_1} \frac{\partial\varphi_1}{\partial N} =
4\psi \frac{1}{\varphi_2} \frac{\partial\varphi_2}{\partial \psi}
+ \psi^2 \frac{1}{\varphi_2} \frac{\partial^2 \varphi_2}{\partial\psi^2}
= - M^2,
\end{equation}
where $M^2$ is a constant of integration. The solution for these
equations are
\begin{eqnarray}
&& \varphi_1(N) = e^{-3N/2} \left[ A_1 e^{\frac{\sqrt{9-4M^2}N}{2}} +
A_2 e^{-\frac{\sqrt{9-4M^2}N}{2}} \right], \\
&& \varphi_2(\psi) = \left(\frac{\psi}{\psi_0}\right)^{-3/2}
\left[ B_1 \left(\frac{\psi}{\psi_0}\right)^{\frac{\sqrt{9-4M^2}}{2}}
+ B_2 \left(\frac{\psi}{\psi_0}\right)^{-\frac{\sqrt{9-4M^2}}{2}} \right],
\end{eqnarray}
where ($A_1, A_2, B_1, B_2$) are constants of integration.
From the second equation in (\ref{sep}), we obtain
\begin{equation}
4\psi \frac{\partial\varphi_b}{\partial\psi} + \psi^2 \frac{\partial^2\varphi_b}{
\partial\psi^2} = - M^2\varphi_b,
\end{equation}
which means that the eq. (\ref{l4}) can be written as
\begin{equation}      \label{cons}
\frac{\partial^2\delta\varphi}{\partial N^2} + 3 \frac{\partial
\delta\varphi}{
\partial N} - e^{-2N} \nabla^2_r\delta\varphi + \psi\left[
4\frac{\partial\delta\varphi}{\partial\psi}
+ \psi \frac{\partial^2\delta\varphi}{
\partial\psi^2}\right]
= -4 M^2 \varphi_b \Phi + 2 \frac{\partial\Phi}{\partial N}
\frac{\partial\varphi_b}{\partial N} + 2\psi \frac{\partial
\varphi_b}{\partial \psi} \Phi.
\end{equation}
This equation is very important because shows how the
gauge-invariant metric
fluctuations and the inflaton field fluctuations are related.
However, as we shall see later, the right hand of this equation
could be zero for a constant $\varphi_b$ and $M=0$. This is the case
in a effective 4D de Sitter expansion, which is the subjet of study
in this letter.
Using the fact that $\delta\varphi(N,\vec r,\psi) =
e^{-3N/2} \left({\psi_0\over\psi}\right)^2 \phi(N,\vec r)$ and
$\Phi(N,\vec r,\psi) = \left({\psi\over\psi_0}\right) \chi(N,\vec r)$,
we obtain that the eq. (\ref{cons}) only is consistent for $M=0$.
Hence, if we require that $\varphi_b$ be consistent with (\ref{l5}), the
solution for $\varphi_b$ holds
\begin{equation}
\varphi_b(N,\psi) = \varphi^{(0)}_b,
\end{equation}
where $\varphi^{(0)}_b = A_1 B_1 =\varphi_b(N=0, \psi=\psi_0)$
is a constant.
Hence, the equation (\ref{cons}) becomes
\begin{equation}\label{new}
\frac{\partial^2\phi}{\partial N^2} - e^{-2N} \nabla^2_r\phi -
\left(\psi^2 \frac{\partial^2\phi}{\partial\psi^2} + \frac{1}{4} \phi\right)
= 0,
\end{equation}
where $\phi(N, \vec r)$ can be expanded as
\begin{equation}
\phi(N,\vec r) = \frac{1}{(2\pi)^{3/2}} {\Large\int} d^3k_r {\Large\int}
dk_{\psi} \left[ A_{k_r k_{\psi}} e^{i \vec{k_r}.\vec r}
\Sigma_{k_r k_{\psi}}(N) + c.c.\right].
\end{equation}
Therefore, the $N$-dependent modes $\Sigma_{k_r k_{\psi}}(N)$
comply with the differential equation
\begin{equation}\label{difeq}
\frac{\partial^2 \Sigma_{k_r k_{\psi}}(N)}{\partial N^2}
+ \left[ k^2_r e^{-2N} - \left( k^2_{\psi} \psi^2 + \frac{1}{4}
\right)\right] \Sigma_{k_r k_{\psi}}(N) = 0.
\end{equation}
The general solution for the eq. (\ref{difeq}) is
\begin{equation}\label{sol}
\Sigma_{k_r k_{\psi}}(N) =
\alpha_1 {\cal H}^{(1)}_{\mu}[x(N)] +
\alpha_2 {\cal H}^{(2)}_{\mu}[x(N)],
\end{equation}
where
($\alpha_1,\alpha_2$) are constants of integration, $\mu= {\sqrt{1+
4(k_{\psi}\psi)^2}\over 2}$ and $x(N) = k_r e^{-N}$.

To normalize the solution (\ref{sol}), we can take only the
homogeneous solution, such that for a generalized Bunch-Davies
vacuum\cite{bd},
we obtain $\alpha_1=0$ and $\alpha_2= i \sqrt{{\pi\over 4}}$.

\section{Effective 4D de Sitter expansion}

In this section we shall study the effective 4D
dynamics in a de Sitter expansion
of both, the metric and the inflaton field fluctuations,
which are reciprocally related.

\subsection{Effective 4D metric}

To study the dynamics of the system on an effective
4D de Sitter expansion, we shall consider
the transformation\cite{plb2006}
\begin{equation}\label{trans}
t = \psi_0 N, \qquad R=\psi_0 r, \qquad \psi = \psi.
\end{equation}
With this transformation the
5D background metric (\ref{back}) becomes
$\left(dS^2\right)_b = \left({\psi\over \psi_0}\right)^2
\left[dt^2 - e^{2t/\psi_0} dR^2\right]-d\psi^2$.
This is known as the Ponce de Leon metric\cite{pdlm} and
is an example of the much-studied class of canonical metrics
$dS^2 = \psi^2 g_{\mu\nu} dX^{\mu} dX^{\nu} - d\psi^2$\cite{MLW}.
To obtain an effective 4D dynamics we can
take a foliation $\psi=\psi_0$ in the Ponce de Leon metric 
\begin{equation}\label{desitter}
\left(dS^2\right)_b \rightarrow \left(ds^2\right)_b
= dt^2 - e^{2t/\psi_0} dR^2.
\end{equation}
This effective 4D metric describes a 4D expansion
of a 3D spatially flat, isotropic and homogeneous universe that
expands with a constant Hubble parameter $H=1/\psi_0$ and a 4D
scalar curvature $^{(4)} {\cal R} = 12 H^2$. Furthermore,
the background energy density is given by
\begin{equation}\label{bac}
\rho_b = \frac{3 H^2}{8\pi G}.
\end{equation}
Once we know the modes $\xi_{k_R k_{\psi_0}}(t)=i \sqrt{\pi \over
4H} {\cal H}^{(2)}_{\mu}[y(t)]$ [with
$\mu = {\sqrt{9+16 k^2_{\psi_0}/H^2}\over 2}$ and
$y(t) = (k_R/H) e^{-H t}$]\cite{plb2006}, we can study
the dynamics for the modes $\Sigma_{k_Rk_{\psi_0}}(t)$ for
a de Sitter expansion. If we take into account
the transformations (\ref{trans}), the equation (\ref{difeq}) assume
the form
\begin{equation}\label{dff}
\ddot{\Sigma}_{k_R k_{\psi_0}}(t) +
\left[ e^{-2H t} k^2_R - \left(k^2_{\psi_0} + \frac{H^2}{4}\right)\right]
\Sigma_{k_R k_{\psi_0}}(t)= 0,
\end{equation}
where $4 k^2_{\psi_0}$ plays the role of the squared mass of the
inflaton field in 4D conventional models of inflation. Note that in our
model this mass is geometrically induced by the fifth coordinate.
Moreover, in a more realistic model
(like a power-law expansion of the universe), the equation for the
modes (\ref{dff}) should be inhomogeneous.
The origin of this inhomogeneity should be
in the right hand of the equation (\ref{cons}).

If we normalize the solution of the differential equation
(\ref{dff}), we obtain ${\Sigma}_{k_R k_{\psi_0}}(t)=
i {\sqrt{\pi/H}\over 2} {\cal H}^{(2)}_{\lambda}[y(t)]$,
where $\lambda = {1\over 2} \sqrt{1 + 4 k^2_{\psi_0}/H^2}$.
This solution give us the time dependent modes for the scalar
field fluctuations $\delta\varphi$ related to the modes of
the gauge-invariant scalar metric fluctuations $\Phi$, for an
effective 4D de Sitter expansion of the universe.

\subsection{Energy density fluctuations and spectrum}

In order to make an analysis for the amplitude
of the energy density fluctuations and its spectrum
we must calculate $\left.\delta\rho = \bar{g}^{NN} \delta T_{NN}
\right|_{N=Ht,\psi=\psi_0=1/H}$,
where $\delta T_{NN} = - {1\over 2} \delta g_{NN} \varphi_{,L} \varphi^{,L}$.
After make a semiclassical approximation for $\varphi$, we obtain
\begin{eqnarray}
\delta\rho & = &
2 \Phi \psi^2 \left(\frac{\partial\varphi}{\partial\psi}\right)^2
\nonumber \\
& = &  8 H^2 \Phi \delta\varphi^2
+ 24 H^2\varphi_0  \Phi\delta\varphi,
\end{eqnarray}
where $\varphi_0$ is the effective 4D background inflaton field.
In a de Sitter expansion, this field
is a constant of $t$ so that the slow-roll conditions\cite{slow} for
this field are
guaranteed.

The expectation value for $\delta\rho$ will be
(we assume the commutation relation
$\left[ \Phi,\delta\varphi\right] =0$)
\begin{equation}
\left<\delta\rho\right>
= 24 H^2 \varphi_0  \left<\Phi \delta\varphi
\right>,
\end{equation}
where $\left<\Phi\delta\varphi\right> = {e^{-3H t}\over 2\pi^2} {\Large\int}
dk_R k^2_R \Sigma_{k_R k_{\psi_0}}(t) \xi^*_{k_R k_{\psi_0}}(t)$.
On the infrared sector (i.e., for $k_R \ll k_H =H e^{H t}$), we obtain
\begin{equation}\label{deltarho}
\left<\delta\rho\right> \simeq \frac{8 G}{\pi^2} H \varphi_0 
2^{\mu+\lambda} \Gamma(\lambda) \Gamma(\mu)\frac{\epsilon^{3-\lambda-\mu}
}{3-\mu-\lambda},
\end{equation}
where $\epsilon ={k^{(IR)}_{max}\over k_p} \ll 1$ is a dimensionless
parameter. Here, $k^{(IR)}_{max} = H e^{H t_i}$, is the
wavenumber related to the Hubble radius at the moment $t_i$ (the time
when the horizon enters) and $k_p$ is the
Planckian wavenumber. If fact, we choose $k_p$ as a cut-off scale
of all the spectrum.
Note that (\ref{deltarho}) takes into account both, the metric
and inflaton fluctuations during an effective 4D de Sitter expansion.
The relative amplitude for the energy density fluctuations
are $\left<\delta\rho\right>/\rho_b$, where $\rho_b$
is the
background energy density for a de Sitter expansion given
by the eq. (\ref{bac}).
To analyze the
spectrum of
$\left<\delta\rho\right>/\rho_b$ on cosmological scales
we can make use of
the result obtained in a previous work\cite{plb2006}. From
the experimental data\cite{ex}, for the energy perturbation
spectral index $n_s=4-\sqrt{9+16 k^2_{\psi_0}/H^2} =0.97\pm 0.03$,
we obtain
\begin{equation}\label{condicion}
0 < \frac{k_{\psi_0}}{H} < 0.15.
\end{equation}
This result was obtained from the spectrum of $\left<\Phi^2\right>$
for a de Sitter expansion and is analogous to
the obtained from the requirement of scale invariance
for a de Sitter expansion in the
standard 4D inflationary models: $\left({m\over H}\right)^2 \ll 1$.

The spectral index $n_1=3-\mu-\lambda$ for
$\left<\Phi\delta\varphi\right>$
in (\ref{deltarho}), is
\begin{equation}  \label{indices}
0.9483 < n_1 < 1,
\end{equation}
which is almost scale invariant as the spectrum of $\left<\Phi^2\right>$.
In a more general model like power-law inflation (where the
background field $\varphi_0$ depends on the time), one
would expect other two spectral indices $n_2$ and $n_3$, which,
could not be nearly scale invariant. This case will be subject of
study in a forthcoming work.

In the figure (1) was plotted
$\left<\delta\rho\right>/\rho_b$ as a function of $\epsilon$
for $k_{\psi_0}/H=0.15$ (continuous line) and
$k_{\psi_0}/H=0.00015$ (pointed line), respectively.
The $\epsilon$-values here plotted corresponds
to $10^4$ or $10^3$ times the typical galactic size today (called
cosmological scales).
Note that in both cases
$\left<\delta\rho\right>/\rho_b$ increases with $\epsilon$ (more
exactly, increases as $\epsilon^{3-\lambda-\mu}$),
and
$\left.\left<\delta\rho\right>/\rho_b\right|_{k_{\psi_0}/H=0.15} \simeq
41.8 \left({\varphi_0\over H}\right) \epsilon^{0.97}$
always is greater (and its slope, too) than
$\left.\left<\delta\rho\right>/\rho_b\right|_{k_{\psi_0}/H=0.00015}
\simeq 42.7 \left({\varphi_0\over H}\right) \epsilon^{0.99}$,
because $\epsilon \ll 1$.
The discrepancy between them increases on smallest scales,
which is a manifestation of how the universe becomes more and more
inhomogeneous as we approach to astrophysical scales.
This is not surprising, because is well known that
$\left<\delta\rho\right>/\rho_b$ on astrophysical scales is
bigger than on cosmological scales.

\section{Final comments}

In this paper we have studied the dynamics
of 4D gauge-invariant metric fluctuations (which are reciprocally
related to inflaton field fluctuations), from a 5D vacuum state.
This topic was examined also in \cite{brand}, but in a rather
different setting.
In particular, we have examined these
fluctuations in an effective 4D de Sitter expansion for the universe
using a first-order expansion for the metric tensor.
It was done in a previous work, but in a partial manner.
In the complete approach here developed, we require that
the spectrum of $\left<\Phi^2\right>$ be
nearly scale invariant, in agreement with the experimental
data\cite{ex}. Once the cut (\ref{condicion}) is obtained,
it is possible to find the power of the spectrum
for $\left<\delta\rho\right>/\rho_b$ [see the cut for $n_1$
in eq. (\ref{indices})].
We obtain that
\begin{displaymath}
\frac{\left<\delta\rho\right>}{\rho_b} \simeq
\frac{64}{3\pi} 2^{\mu+\lambda} \frac{\varphi_0}{H}
\Gamma(\lambda) \Gamma(\mu)
\frac{\epsilon^{3-\mu-\lambda}}{(3-\mu-\lambda)},
\end{displaymath}
take values
of the order of $10^{-5}$ for $H \simeq 10^{-9} \  M_p$ and
$\varphi_0 \simeq 10^{-12} \  M_p$, so that the model provides
a sufficient
number of e-folds (i.e., $N_e >60$) required to solve the
horizon/flatness problem for $t > 6 \times 10^{10} \  G^{1/2}$.
It is very important that $\varphi_0$ can take
sub-Planckian values. It solves one of the problems
of standard 4D chaotic inflation, in which the scalar field remains always
with trans-Planckian values.
The consequences
of this complete approach are very important, because we find
that the spectrum of $\left<\delta\rho\right>/\rho_b$ now
depends on the coupling of $\Phi$ with $\delta\varphi$.
Moreover, this spectrum is almost scale invariant
(as the spectrum of $\left<\Phi^2\right>$), for $k_{\psi_0}/H \ll 1$.
However, one would expects additional spectral indices in a more
realistic inflationary model where $\varphi_0$ depends on the time.
This case will be studied in a forthcoming work. \\
\vskip .3cm
\noindent
{\bf Acknowledgements}\\
\noindent
The authors acknowledge R. Brandenberger for interesting comments.
A. M. acknowledges UNMdP for financial support.
M.B. acknowledges CONICET and UNMdP (Argentina) for financial
support.\\

\newpage
\begin{displaymath}
\end{displaymath}

\vskip 7cm
\noindent
Fig. 1) Evolution of $\left<\delta\rho\right>/\rho_b$
as a function of $\epsilon$ for
$k_{\psi_0}/H =0.15$ (continuous line) and
$k_{\psi_0}/H =0.00015$ (pointed line).\\

\end{document}